\documentclass{appolb}

\usepackage{amsmath,amsfonts,amssymb}
\usepackage[dvips]{graphicx}
\usepackage{multicol}

\begin{document}

\title{On Capital Dependent Dynamics of Knowledge 
\thanks{The paper was presented at 2nd Polish Symposium on Econo- and Sociophysics, 
Krakow, Poland, 21-22 April 2006. The authors acknowledge support of the Rector's 
Scholarship Fund 2004/2005.}}

\author{Marek Szyd{\l}owski
\address{Astronomical Observatory, Jagiellonian University \\
Orla 171, 30-244 Krak{\'o}w, Poland \\
M. Kac Complex Systems Research Centre, Jagiellonian University \\ 
Reymonta 4, 30-059 Krak{\'o}w, Poland \\
\texttt{uoszydlo@cyf-kr.edu.pl}}
\and
Adam Krawiec 
\address{Institute of Public Affairs, Jagiellonian University \\ 
Rynek G{\l}{\'o}wny 8, 31-042 Krak{\'o}w, Poland \\
M. Kac Complex Systems Research Centre, Jagiellonian University \\ 
Reymonta 4, 30-059 Krak{\'o}w, Poland \\
\texttt{uukrawie@cyf-kr.edu.pl}}
}

\maketitle

\begin{abstract}
We investigate the dynamics of growth models in terms of dynamical 
system theory. We analyse some forms of knowledge and its influence 
on economic growth. We assume that the rate of change of knowledge 
depends on both the rate of change of physical and human capital. 
First, we study model with constant savings. The model with 
optimised behaviour of households is also considered. We show that 
the model where the rate of change of knowledge depends only on 
the rate of change of physical capital can be reduced to the form 
of the two-dimensional autonomous dynamical system. All possible 
evolutional paths and the stability of solutions in the phase space 
are discussed in details. We obtain that the rate of growth of 
capital, consumption and output are greater in the case of 
capital dependent rate of change of knowledge. 
\end{abstract}


\section{Introduction}

In his model Solow \cite{Solow56} introduced capital, labour and `knowledge' 
as the most important inputs which are used to produce output. Knowledge 
can be everything else apart from capital and labour and can play the role 
of technological progress. However, the growth of knowledge was exogenous. 
There are many attempts to describe how knowledge affects output 
\cite{Romer96}. There are different ways to include knowledge as an 
endogenous variable in a model. For example, knowledge can be produced in 
research and development sector \cite{Romer90}. Knowledge can also be 
treated as another kind of input, human capital, used in production 
\cite{Mankiw92}. 

We propose that the change of physical capital influences the change of 
knowledge. We assume that the rate of knowledge growth is proportional to 
the rate of capital growth. It can be interpreted in different ways.
We can think that capital has some positive externalities on technological 
progress. Another possibility is that some capital is used directly in 
research and development, for example, it could be supercomputers, satellites 
or other equipment. 

In this paper we study the dynamics of the optimal growth model with such 
a kind of endogenous technological progress. We compare this model with the 
optimal growth model with exogenous knowledge. We find how much the economy  
with endogenous knowledge grows faster than the economy with exogenous 
knowledge for different values of model parameters.

\section{Capital dependent model of growth of knowledge}

We consider the economy where output $Y$ is produced by using 
physical capital $K$, human capital $H$, labour $L$, and knowledge $A$ 
as inputs 
\begin{equation}
\label{eq:1}
Y(t) = F(K(t), H(t), A(t)L(t)).
\end{equation}
This production function has constant returns to scale in $K(t)$, $H(t)$, and 
$A(t)L(t)$. Labour and knowledge enter multiplicatively to the production 
function, and $A(t)L(t)$ is also called effective labour. We assume that 
labour increases in the constant rate $n$ 
\begin{equation}
\label{eq:2}
\frac{\dot{L}}{L} =n
\end{equation}
where an overdot means the differentiation with respect to time. 

The neoclassical model of economic growth is based on simplified 
assumption that knowledge grows with the constant rate. There are 
some propositions of relaxing this assumption. We also propose 
some alternative modification of exponential growth of knowledge. 
Our idea is to consider the more general assumption by including 
the capital. We assume that apart from the exogenous growth of 
knowledge both physical and human capital can influence on 
the rate of growth of knowledge. We assume that these processes 
are additive and proportional to rates of growth of these capitals 
\begin{equation}
\label{eq:51}
\frac{\dot{A}}{A} = g + \mu \frac{\dot{K}}{K} + \nu \frac{\dot{H}}{H}
\end{equation}
or 
\begin{equation}
\label{eq:52}
A = A_0 e^{gt} K^{\mu} H^{\nu}.
\end{equation}
For $\mu=\nu=0$ we obtain the constant rate of growth of knowledge. 
The interpretation of the above assumption can be following. 
The physical capital is necessary in research of scientific and 
industrial laboratories. It is especially important in contemporary 
science. 




Let us apply the dynamical systems methods \cite{Lorenz89} to the model of 
growth with capital dependent growth of rate of knowledge. 
\begin{subequations}
\label{eq:53}
\begin{align}
\dot{k} &= (1-\mu)s_{k} k^{\alpha} h^{\beta} 
- \nu s_{h} k^{\alpha+1} h^{\beta-1} 
- [(1-\mu-\nu)\delta + n +g]k \\
\dot{h} &= (1-\nu)s_{h} k^{\alpha} h^{\beta} 
- \mu s_{k} k^{\alpha-1} h^{\beta+1} 
- [(1-\mu-\nu)\delta + n +g]h 
\end{align}
\end{subequations}

\begin{figure}
\centering
\includegraphics[angle=-90,width=\textwidth]{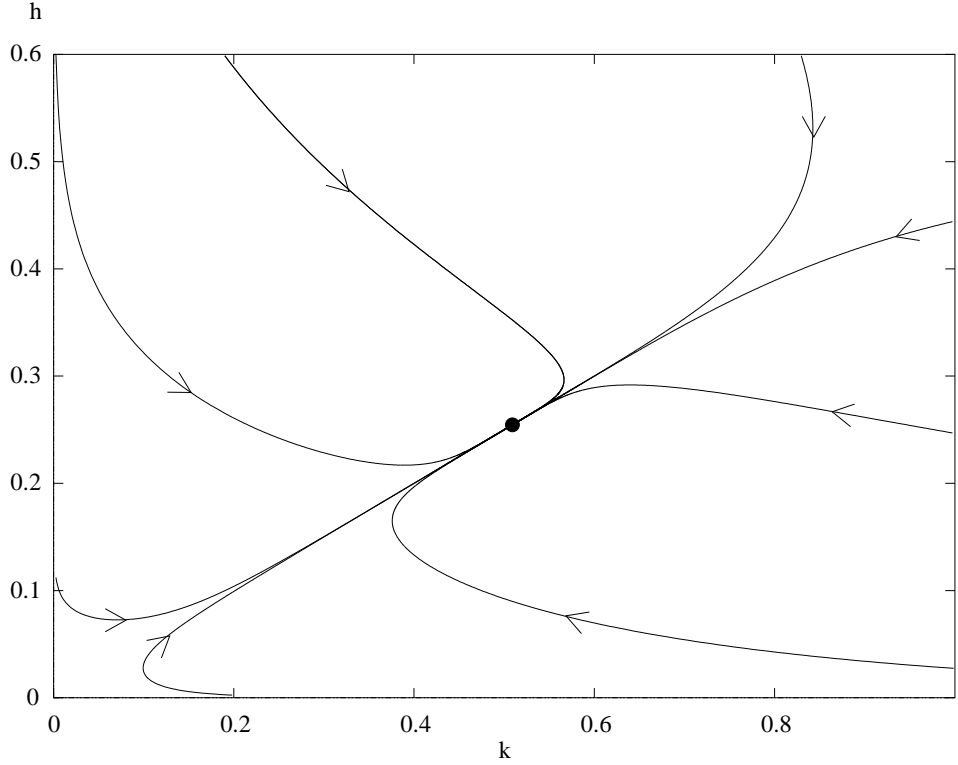}
\caption{The phase portrait for system~(\ref{eq:51})}
\label{fig:0}
\end{figure}

System (\ref{eq:51}) have at least two critical points in finite domain 
of phase space. In the Fig.~\ref{fig:0} we choose for presentation 
$\delta=0.007$, $\mu=0.2$, $\nu=0.2$, $\alpha=0.35$, $\beta=0.4$, 
$n=0.02$, $g=0.04$. The critical point located at the origin is 
a saddle while the second one is a stable node. For different values 
of the parameters the node is located on the the line $k \propto h$. 

The comparison of the phase portraits for $\mu\ne0$, $\nu\ne0$ with 
$\mu=\nu=0$ gives that while they are topologically equivalent 
the node for the latter case is located for higher $k$, $h$.

\subsection{Optimisation in the Model with Endogenous Knowledge}

At first we avoid to explore the nature of knowledge and take the simplest 
assumption that knowledge has the exogenous character and grows in the 
constant rate $g$,
\begin{equation}
\label{eq:3}
\frac{\dot{A}}{A} = g.
\end{equation}

The capital accumulation comes from output which is not consumed. Taking 
into account the capital depreciation $\delta$, capital change is given by
\begin{equation}
\label{eq:4}
\dot{K} = F(K(t),A(t)L(t)) - C(t) - \delta K(t). 
\end{equation}
It is convenient to use the variables in units of effective labour $AL$ 
(denoted in small letters). In this case we obtain 
\begin{equation}
\label{eq:5}
\dot{k} = f(k(t)) -c - (g+n+\delta)k(t).
\end{equation}

In the original Solow model the savings are a fixed share of product. 
However, we can allow the households to choose between saving and 
consumption in their lifetime \cite{Ramsey28}. It means that the infinitely 
living households such a level of consumption over time to maximise their 
utility function 
\begin{equation}
\label{eq:6}
U = \int_{0}^{\infty} e^{-\rho t} u(C(t)) dt
\end{equation}
where $\rho$ is a discount rate. 

To solve the maximisation problem we use the Pontryagin Maximum Principle 
\cite{Dixit90}. As the result we obtain the system of two differential 
equations 
\begin{subequations}
\label{eq:7}
\begin{align}
\dot{k} &= k^{\alpha} - c - (\delta + g + n)k \\
\dot{c} &= \frac{c}{\sigma} (\alpha k^{\alpha-1} - \delta - g - n -\rho).
\end{align}
\end{subequations}
To obtain this system we assume the Cobb-Douglas production function
\[
f(k) = k^{\alpha}
\]
as well as the constant-relative-risk-aversion (CRRA) utility function
\begin{equation}
\label{eq:8}
u(c(t)) = \frac{c(t)^{1-\sigma}}{1-\sigma}.
\end{equation}
which is characterised by the constant elasticity of substitution between 
consumption in any two moments of time.

Let us consider the dynamics of system~(\ref{eq:5}). For simplification 
we put $b_{1}=\delta + g + n$ and find three critical points: \\
the unstable node 
\begin{equation}
k_1 = c_1 = 0,
\end{equation}
the stable node
\begin{equation}
k_2 = b_{1}^{1/(\alpha-1)}, \quad c_2 = 0,
\end{equation}
and the saddle 
\begin{subequations}
\label{eq:p1c}
\begin{align}
k_3 =& \left(\frac{b_{1} +\rho}{\alpha}\right)^{1/(\alpha-1)} \\
c_3 =& \left(\frac{b_{1} +\rho}{\alpha}\right)^{\alpha/(\alpha-1)}
- b_{1}
\left(\frac{b_{1} +\rho}{\alpha}\right)^{1/(\alpha-1)}
\end{align}
\end{subequations}
Two first points have no economic concern because they represent economies 
without consumption. Only the third critical point, the saddle, is relevant 
in our discussion. Households choose such a level of consumption which is 
optimal for a given amount of capital. And it always lies on one of two 
trajectories moving to the saddle point. Once the economy reach the saddle 
point it enters the balanced growth path where all quantities per a unit of 
effective labour are constant. However, capital, consumption, and output as 
well as their counterparts per capita (per unit of labour alone) increase in 
time. 

\begin{figure}
\centering
\includegraphics[angle=-90,width=\textwidth]{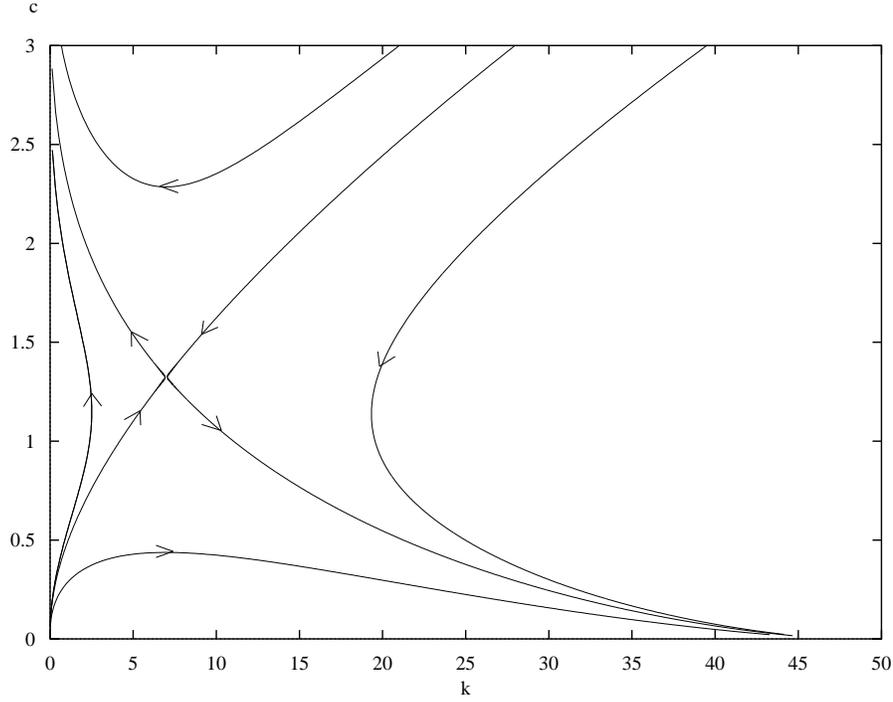}
\caption{The phase portrait of system~(\ref{eq:7})}
\label{fig:1}
\end{figure}

The phase portrait of this system is shown on Fig.~\ref{fig:1}. The bold 
lines denote two trajectories which lead to the saddle.  

Let us return now to the endogenous technological progress. However, we 
consider that only the rate of growth of physical capital has influence on 
knowledge. Then equation~(\ref{eq:51}) assumes the form 
\begin{equation}
\label{eq:10}
\frac{\dot{A}}{A} = g + \mu \frac{\dot{K}}{K}.
\end{equation}
We assume that some part of technological progress has the exogenous character. 
There is also the additional term which describes the influence of change in 
capital stock on the knowledge growth. The proportionality parameter $\mu$ 
belongs to $[0,1)$. For $\mu=0$ we have the model with the exogenous knowledge 
analysed in the previous section. This additional component could be 
interpreted as the capital equipment used in research and development.  

Assuming the form of the production function and the utility function as in 
the previous section, the optimisation procedure gives us the following 
two-dimensional dynamical system 
\begin{subequations}
\label{eq:11}
\begin{align}
\dot{k} &= (1-\mu) k^{\alpha} - (1-\mu) c - 
[(1-\mu)\delta + g + n]k \\
\dot{c} &= \frac{c}{\sigma} [\alpha (1-\mu)k^{\alpha-1} 
- (1-\mu)\delta - g - n -\rho].
\end{align}
\end{subequations}
When we put $\mu = 0$ we obtain system~(\ref{eq:7}). We use this feature to 
compare the dynamics of system~(\ref{eq:11}) with system~(\ref{eq:7}).

\begin{figure}
\centering
\includegraphics[angle=-90,width=\textwidth]{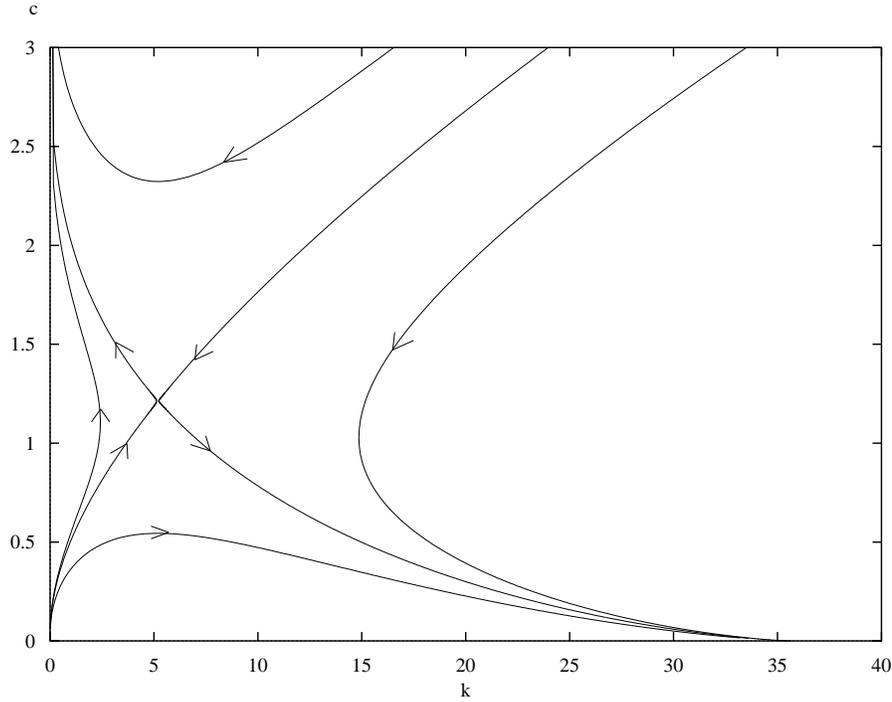}
\caption{The phase portrait of system~(\ref{eq:11})}
\label{fig:2}
\end{figure}

For simplification we denote $b_{2}=(1-\mu)\delta + g + n$. 
System~(\ref{eq:11}) has three critical points: \\
the unstable node 
\begin{equation}
k_1 = c_1 = 0,
\end{equation}
the stable node
\begin{equation}
k_2 = \left( \frac{b_{2}}{1-\mu} \right)^{1/(\alpha-1)}, 
\quad c_2 = 0,
\end{equation}
and the saddle 
\begin{subequations}
\label{eq:p2c}
\begin{align}
k_3 =& \left(\frac{b_2 + \rho}{\alpha (1-\mu)}\right)^{1/(\alpha-1)} \\
c_3 =& \left(\frac{b_2 + \rho}{\alpha (1-\mu)}\right)^{\alpha/(\alpha-1)}
- \frac{b_2}{1-\mu} 
\left(\frac{b_2 + \rho}{\alpha (1-\mu)}\right)^{1/(\alpha-1)}
\end{align}
\end{subequations}
The dynamics of system~(\ref{eq:11}) is presented on Fig.~\ref{fig:2}. 
Comparing with Fig.~\ref{fig:1} we can see that both phase portraits are 
topologically equivalent. The systems are structurally stable. 

Two nodes represent unrealistic economies with zero level consumption. The 
households choose the optimal levels of consumption for given capital 
stock. These choices forms two trajectories which approach the saddle 
solution. When the economy converges to the saddle point it reaches the 
balanced growth path. Capital, consumption, and output per unit of effective 
labour are constant. The dynamics of capital, consumption, output and capital, 
consumption, output per a unit of labour depend on the parameters $g$, $n$, 
and $\mu$. 

Table~\ref{tab:1} presents the rates of change of capital, consumption, 
output as well as their per capita counterparts in both considered models. 
{\centering
\begin{table}
\caption{The rates of growth of capital $K$, consumption $C$, and output $Y$ 
in the models with exogenous and endogenous growth of knowledge}
\label{tab:1}
\begin{tabular}{lcc}
\hline
\rule{0pt}{4ex} \ variables \ \ & rate of growth with & rate of growth with \\
& \ \ exogenous knowledge \ \ & \ \ endogenous knowledge \ \\
\hline \hline
\rule{0pt}{4ex} $K$, $C$, $Y$ & $g+n$ & $\frac{g+n}{1-\mu}$ \\
\rule{0pt}{4ex} $\frac{K}{L}$, $\frac{C}{L}$, $\frac{Y}{L}$ & $g$ 
& $\frac{g+\mu n}{1-\mu}$ \\ & & \\
\hline
\end{tabular}
\end{table}
}

\begin{figure}
\centering
\includegraphics[width=0.8\textwidth]{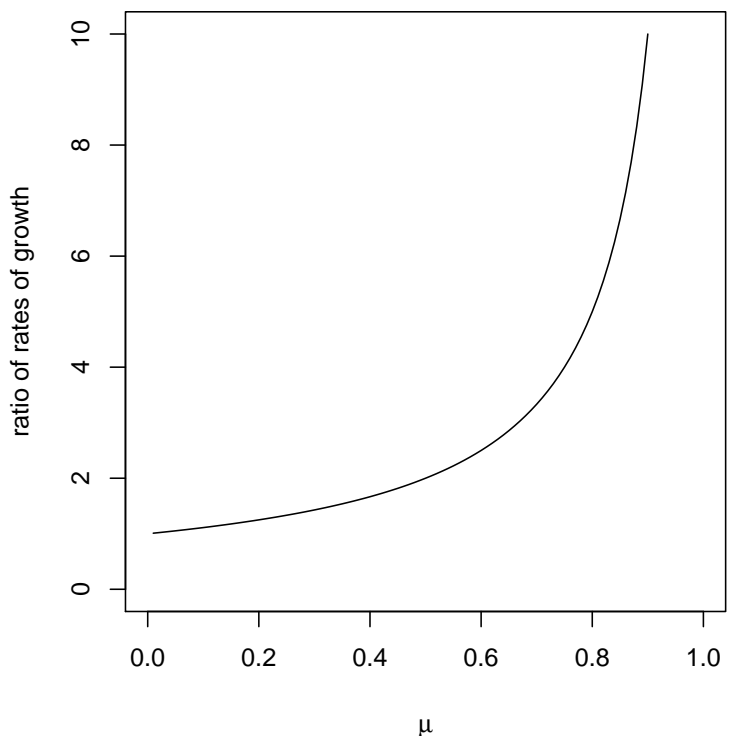}
\caption{The dependence of ratio of rates of growth of capital, consumption, 
output in the models with endogenous and exogenous knowledge on the 
parameter~$\mu$}
\label{fig:3}
\end{figure}

We can compare the rate of growth of capital, consumption, and output 
in the models with endogenous and exogenous knowledge. The ratio of 
rates of growth for capital, consumption, and output is 
\begin{equation}
\label{eq:20}
R_{X} = \frac{\frac{g+n}{1-\mu}}{g+n} = \frac{1}{1-\mu},
\end{equation}
where $X$ means $K$, $C$, and $Y$. The ratio of rates of growth of capital, 
consumption, output in these two models depends only on the parameter $\mu$. 
The rate of growth of all the three variables is greater in the presence of 
endogenous knowledge. Figure~\ref{fig:3} shows how many times the rate of 
growth in the model with endogenous knowledge is greater than in the model 
with exogenous knowledge for different values of $\mu$. For example, for 
$\mu = 0.2$ the rate of growth is 25\% higher, and for $\mu = 0.5$ the rate 
of growth is 2 times higher, in the model with endogenous technological 
progress than in the model model with exogenous technological progress. 

\begin{figure}
\centering
\includegraphics[width=0.8\textwidth]{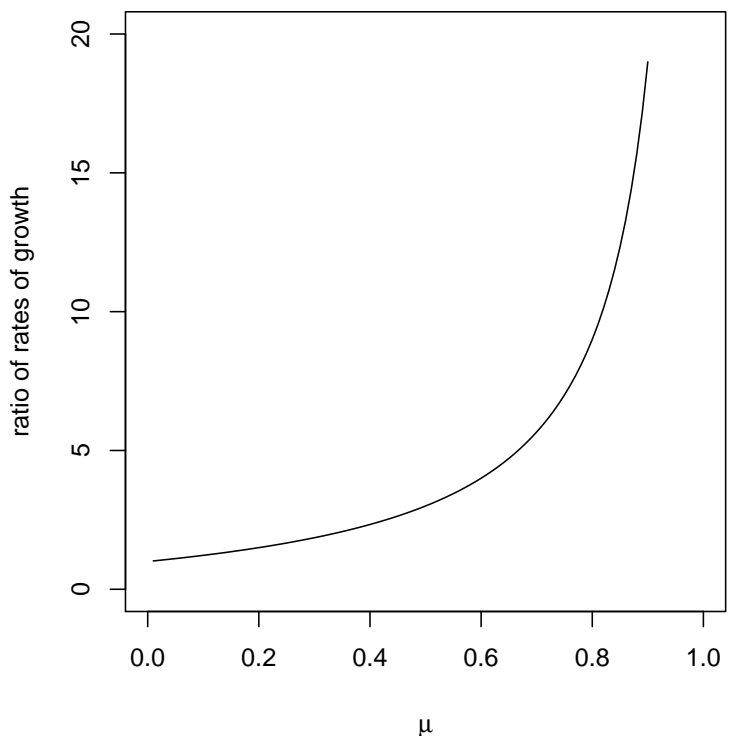}
\caption{The dependence of ratio of rates of growth of capital, consumption, 
output per unit of labour in the models with endogenous and exogenous 
knowledge on the parameter $\mu$.}
\label{fig:4}
\end{figure}

We can also compare the rates of growth of per capita quantities in the models 
with endogenous and exogenous technological progress. The ratio of rates of 
growth is 
\begin{equation}
\label{eq:21}
R_{X/L} = \frac{\frac{g+\mu n}{1-\mu}}{g} 
= \frac{g + \mu n}{g(1-\mu)}.
\end{equation}
The ratio of rates of growth of capital, consumption, output per unit of 
labour in these two models depends both on the parameter $\mu$ and $g$. 
Fig.~\ref{fig:4} presents the ratio of rates of growth with respect to the 
parameter $\mu$. For example assuming the same values of parameters $g$ 
and $n$ we can find that for $\mu = 1/3$ the rate of growth is 2 times 
higher, and for $\mu = 2/3$ the rate of growth is 5 times higher, in the 
model with endogenous technological progress than in the model model with 
exogenous technological progress. When $g>n$ ($g<n$) the ratio is lower 
(higher) for a given $\mu$.  

\subsection{Conclusions} 

We investigate the dynamics of growth models in terms of dynamical 
system theory. We analyse some forms of knowledge and its influence 
on economic growth. We assume that the rate of change of knowledge 
depends on both the rate of change of physical and human capital. 
First, we study model with constant savings. The model with 
optimised behaviour of households is also considered. We show that 
the model where the rate of change of knowledge depends only on 
the rate of change of physical capital can be reduced to the form 
of the two-dimensional autonomous dynamical system. All possible 
evolutional paths and the stability of solutions in the phase space 
are discussed in details. We obtain that the rate of growth of 
capital, consumption and output are greater in the case of 
capital dependent rate of change of knowledge. 

Our proposition of parameterisation of knowledge seems to be a unification 
of exogenous and endogenous factors. If we consider three different 
cases of endogenous ($g=0$, and $\mu \ne 0$ or $\nu \ne 0$) 
exogenous ($g \ne 0$, and $\mu = \nu = 0$) and mixed 
($g \ne 0$, and $\mu \ne 0$ or $\nu \ne 0$), we find that the 
qualitative dynamics is the same for reasonable values of the rest parameters
of the model. The only observable difference is different values of 
rates of change of the phase variables at the critical point. In other words 
the endogenous factors give additional contribution to the rate of change 
of the variables. 

We presented the modification of the Ramsey model of optimal economic growth 
where knowledge growth depends on the rate of growth of physical capital. 
We compare this model with the optimal growth model with the constant rate of 
growth of knowledge. 

We reduced the growth model with physical capital dependence of knowledge 
to two-dimensional dynamical system and investigated its solutions using the 
qualitative methods of dynamical systems. We presented the dynamics of the
models on the phase portraits. 

We calculated the rates of growth of capital, consumption, 
and output as well as their counterparts per capita. We compared these 
rates for both models and found how many times faster the model variables 
grows in the model with endogenous knowledge than in the model with 
exogenous knowledge.

It can be interpreted that physical capital growth add to the rate 
the knowledge growth some additional impact which makes the 
physical capital, consumption and output to grow faster.


\end{document}